# WISDOM OF THE CONFIDENT: USING SOCIAL INTERACTIONS TO ELIMINATE THE BIAS IN WISDOM OF THE CROWDS

GABRIEL MADIROLAS AND GONZALO G. DE POLAVIEJA, Cajal Institute, CSIC, Spain

## 1. INTRODUCTION

Human groups can perform extraordinary accurate estimations compared to individuals by simply using the mean, median or geometric mean of the individual estimations [Galton 1907, Surowiecki 2005, Page 2008]. However, this is true only for some tasks and in general these collective estimations show strong biases. The method fails also when allowing for social interactions, which makes the collective estimation worse as individuals tend to converge to the biased result [Lorenz et al. 2011]. Here we show that there is a bright side of this apparently negative impact of social interactions into collective intelligence. We found that some individuals resist the social influence and, when using the median of this subgroup, we can eliminate the bias of the wisdom of the full crowd. To find this subgroup of individuals more confident in their private estimations than in the social influence, we model individuals as estimators that combine private and social information with different relative weights [Perez-Escudero & de Polavieja 2011, Arganda et al. 2012]. We then computed the geometric mean for increasingly smaller groups by eliminating those using in their estimations higher values of the social influence weight. The trend obtained in this procedure gives unbiased results, in contrast to the simpler method of computing the median of the complete group. Our results show that, while a simple operation like the mean, median or geometric mean of a group may not allow groups to make good estimations, a more complex operation taking into account individuality in the social dynamics can lead to a better collective intelligence.

## 2. RESULTS

To test the wisdom of the confident, we analyzed wisdom of the crowd experiments that studied the impact of social interactions [Lorenz et al. 2011]. In these experiments, groups of 12 subjects were asked to make five consecutive numerical estimations about six different geographical or social facts, v. gr. the border length between Italy and Switzerland, or number of murders in Zurich the previous year. Between trials, subjects in four groups received the arithmetic mean of the 12 estimates of the previous trials, while subjects in other four groups received a diagram with all the estimations of the group in all previous trials. Four control groups received no information between trials. We focus only in the shift of the estimations between trial 1, when no social information had already been provided, and trial 2, when social information had been provided for the first time.
We modelled humans as imperfect estimators that combine private and public information with different relative weights. This is a model we have used before in collective decision-making in fish and ant groups Perez-Escudero & de Polavieja 2011, Arganda et al. 2012]. Here we adapt this modelling approach to the case of human data [Lorenz et al. 2011], in which individuals estimate quantities that can take any positive real number. The probability distribution of answers given by humans based on their private information alone, $p$, is a log-normal distribution,

2   G. Madirolas and G. G. de Polavieja

$$f_X(x|p) = \frac{1}{x\sigma_p\sqrt{2\pi}} e^{-\frac{1}{2}\left(\frac{\log(x)-\mu_p}{\sigma_p}\right)^2}, \tag{1}$$

with median $x_p \equiv \exp(\mu_p)$ a good indicator of the collective knowledge and $\sigma_p$ a measure of the diversity (Figure 1, blue). Modelling humans as estimators that combine private information $p$ and social information $s$ [Pérez-Escudero & de Polavieja 2011], we obtain that the distribution of answers after social influence is of the form

$$f_X(x|p,s) = \frac{1}{x\sigma_f\sqrt{2\pi}} e^{-\frac{1}{2}\left(\frac{\log(x)-\mu_f}{\sigma_f}\right)^2}, \tag{2}$$

again a log-normal distribution, but with median now combination of the private median and the social information, $\mu_f = w_p\mu_p + w_s\mu_s$, with $w_p$ and $w_s$ the private and social weights. If subjects are provided all the estimations given by others, then there is no shift in the median as $\mu_s = \mu_p$ but there is a shift to higher values if the arithmetic mean is provided, as also seen in the data (Figure 1A,B, red). After the social influence, the distribution has a smaller standard deviation, $\sigma_f = \sqrt{1-w_s}\,\sigma_p$, the smaller the higher the social weight (Figure 1A,B, red).

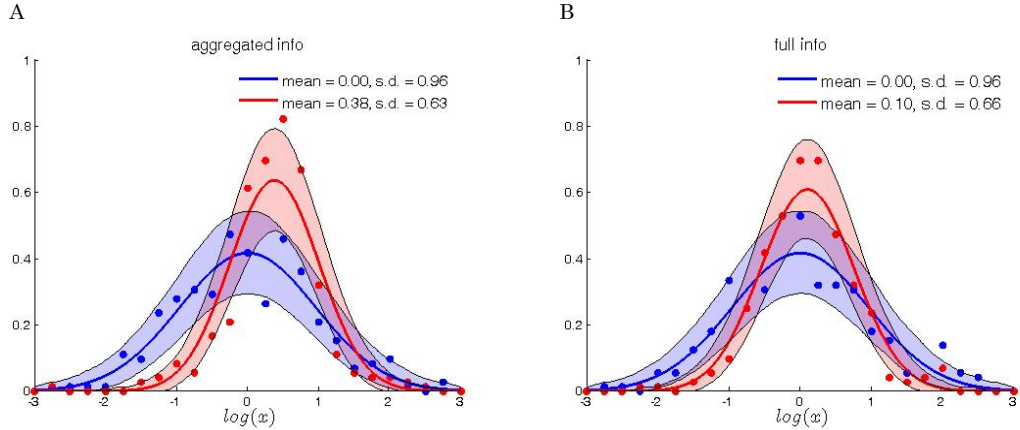

Fig. 1 Probability distribution of estimations with private information only (blue) and after receiving social influence (red). To plot together estimations about six different questions, the variable is standardized by subtracting to the logarithm the mean of the logarithm of estimations at the specific question and then dividing by the standard deviation of them. The points are the experimental frequencies of intervals of width 0.2. Solid line is a Gaussian fit of the frequencies. Shadowed surface is the area where is 90 per cent of the experiments expected to be found given the theoretical fit. The social influence was either by giving to each individual the mean of the other individuals (A) or all the individual answers (B).

Suggested by (2), we expressed the integration made by a subject of her first estimation $x_1$ and the social information provided to her, $\mu_s$, to produce a second estimation $x_2$ as

$$\log(x_2) = w_p\log(x_1) + w_s\mu_s. \tag{3}$$

That is, individuals would be combining private and social information in the logarithmic domain. The private and social confidence weights were assumed to be the same for all individuals in the statistical analysis of Figure 1, but this needs not be the case. To obtain whether there is a relevant individuality in how humans respond to social influence, we computed the social weight from Eq. (3) for each individual

$$w_s = \frac{\log(x_2) - \log(x_1)}{\mu_s - \log(x_1)}. \tag{4}$$

The distribution of these values shows a structure indicating individual differences (Figure 2A). Some individuals are not influenced socially (peak at $w_s = 0$ in Figure 2A), others do not use their private information (peak at $w_s = 1$ in Figure 2A), the majority have intermediate values, and even some others shift their values to values higher (lower) than the social (private) value.

We are interested in isolating those members confident in their private values, that is, those with values of $w_s$ close to 0. We systematically eliminated subjects for which $|w_s| > \omega$, with $\omega$ increasingly small positive real numbers. Then, for the selected individuals, we computed the geometric mean of their first estimation as an indicator of the wisdom of the confident (Fig. 2B for the estimation of the length of the Swiss-Italian border).

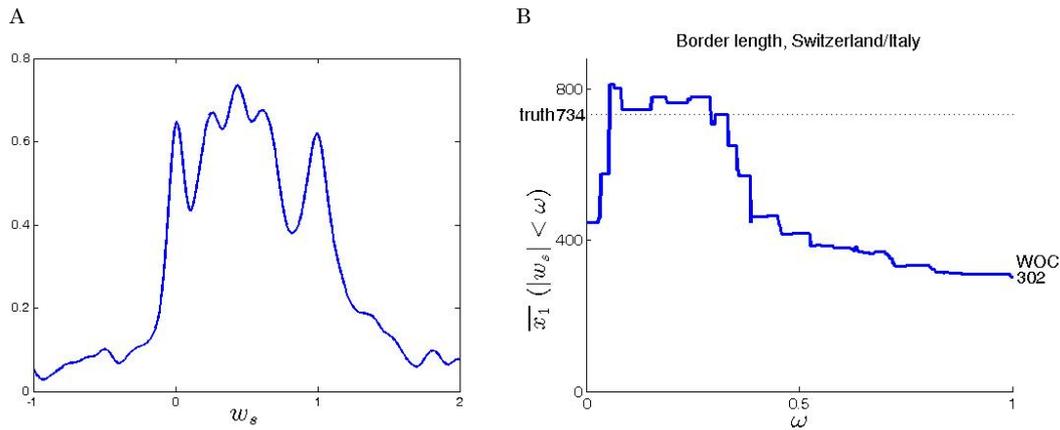

Fig. 2. Wisdom of the confident. Distribution of social weights $w_s$ (A) and Geometric mean $\bar{x}_1$ of estimates made in first trial by those individuals who apply a social weight smaller than some value $\omega$ (B). The question was 'What is the length of the border between Switzerland and Italy in kilometres', and the correct answer is 734. The wisdom of the crowd value 302. Data from [Lorenz *et al.* 2011]